\documentclass[10pt,journal]{IEEEtran}
\IEEEoverridecommandlockouts

\usepackage{graphicx}
\usepackage{amsmath,amsthm,amsfonts,amssymb}
\usepackage{cite,hyperref}
\usepackage{bm}
\usepackage{bbm}
\usepackage{url}
\usepackage{array}
\usepackage{color,soul}
\usepackage{multirow}
\usepackage{booktabs}
\usepackage[table,xcdraw]{xcolor}
\usepackage{enumitem}
\usepackage{subcaption}

\usepackage[english]{babel}
\usepackage{algorithm}
\usepackage{algorithmic}
\usepackage{setspace}
\usepackage[english]{babel}
\usepackage{makecell}

\theoremstyle{plain}

\allowdisplaybreaks[4]

\begin{document}
\title{Rich-ARQ: From 1-bit Acknowledgment to Rich Neural Coded Feedback}

\author{
Enhao Chen and Yulin Shao
\thanks{The authors are with the Department of Electrical and Electronic Engineering, The University of Hong Kong, Hong Kong S.A.R. (e-mails: \url{aruleuer@gmail.com}, \url{ylshao@hku.hk}).
}
}

\maketitle

\begin{abstract}
This paper reimagines the foundational feedback mechanism in wireless communication, transforming the prevailing 1-bit binary ACK/NACK with a high-dimensional, information-rich vector to transform passive acknowledgment into an active collaboration. 
We present Rich-ARQ, a paradigm that introduces neural-coded feedback for collaborative physical-layer channel coding between transmitter and receiver. 
To realize this vision in practice, we develop a novel asynchronous feedback code that eliminates stalling from feedback delays, adapts dynamically to channel fluctuations, and features a lightweight encoder suitable for on-device deployment. 
We materialize this concept into the first full-stack, standard-compliant software-defined radio prototype, which decouples AI inference from strict radio timing. 
Comprehensive over-the-air experiments demonstrate that Rich-ARQ achieves significant SNR gains over conventional 1-bit hybrid ARQ and remarkable latency reduction over prior learning-based feedback codes, moving the promise of intelligent feedback from theory to a practical, high-performance reality for next-generation networks.
\end{abstract}

\begin{IEEEkeywords}
Rich-ARQ, edge intelligence, feedback coding, ultra-reliable communications, AI-hardware co-design.
\end{IEEEkeywords}

\section{Introduction}
\subsection{Motivation}
Feedback is the cornerstone of stability and coordination in any interactive system, enabling adaptation and ensuring reliable operation \cite{lovefeedback,Polyanskiy,rfc3366,shao2021federated}. In modern communication systems, this principle is crystallized in the nearly ubiquitous use of the 1-bit Acknowledgment (ACK) or Negative Acknowledgment (NACK). From the automatic repeat request (ARQ) \cite{comroe1984arq} in Wi-Fi and Bluetooth to the hybrid ARQ (HARQ) \cite{zhao2005practical} protocols in 4G LTE and 5G NR, this minimal feedback mechanism forms the essential control loop for reliable data delivery, triggering retransmissions or complementary parity bits upon failure.

While immensely successful, this foundational design carries an intrinsic limitation: its poverty of information. The single-bit feedback conveys only a coarse binary outcome, i.e., ``success'' or ``failure'', discarding all nuanced information regarding the nature of channel impairments or the decoder's internal state. This constraint is not a flaw in these protocols, which remain marvels of efficient engineering, but it does establish a fundamental performance boundary. 
As wireless systems push toward the extremes of spectral efficiency, this boundary becomes critically salient for ultra-reliable communication in practical scenarios characterized by time-varying channels and short-packet transmission \cite{polyanskiy2010channel,vucetic2002adaptive}.

This begs a fundamental question: \emph{What if the feedback packet could convey far more than a single bit?} 
Imagine a paradigm where feedback is not a final verdict, but a rich, informative signal: a multi-dimensional vector conveying decoder uncertainty, instantaneous channel state, and other soft information \cite{Shannon,shao2024theory}. This transforms the role of feedback from a passive judge to an active collaborator in the communication process. We term this vision Rich-ARQ: a paradigm shift from 1-bit decision feedback to high-dimensional, coded feedback for collaborative physical-layer channel coding.

Realizing this vision, however, presents a profound challenge. Traditional coding theory lacks the tools to design encoders and decoders that can dynamically synthesize and exploit such rich, high-dimensional feedback. The advent of deep learning (DL) \cite{NatureDL} offers a breakthrough pathway. Their unparalleled ability of deep neural networks (DNNs) to learn complex, non-linear mappings makes them ideal agents to extract rich feedback at the receiver and generate optimally coded responses at the transmitter, a process we call neural coded feedback. 

\subsection{Feedback Channel Coding}
The evolution of feedback channel coding traces a clear path from theoretical constructs under idealized assumptions to data-driven designs grappling with real-world complexity. 

The theoretical promise of feedback was established by Shannon \cite{Shannon}, who demonstrated that while noiseless, causal feedback does not increase the capacity of a memoryless forward channel, it can profoundly enhance reliability in the finite blocklength regime. This insight inspired the development of constructive coding schemes aiming to approach this performance limit.

The most celebrated example is the Schalkwijk-Kailath (SK) scheme \cite{SK1,SK2} for the additive white Gaussian noise (AWGN) channel, which employs a linear coding strategy to achieve a double-exponential decay of error probability with blocklength. This landmark result powerfully demonstrated the transformative potential of well-structured feedback. Subsequent schemes, such as the Horstein scheme for the binary symmetric channel \cite{horstein2003sequential,shayevitz2011optimal}, extended this philosophy to other discrete memoryless channels \cite{li2015efficient}.

However, a critical and often prohibitive limitation of these classical schemes is their acute sensitivity to feedback noise.
To address the vulnerability, the Modulo-SK scheme was introduced \cite{MSK}. This scheme incorporates a key innovation: it uses a modulo operation to ``wrap'' the residual estimation error into a bounded interval before transmission. This nonlinear step prevents the error from growing uncontrollably due to noise in the feedback loop, thereby restoring robustness and stability. The Modulo-SK scheme is a pivotal example of extending classical feedback theory to more practical, non-ideal conditions.

The advent of DL provided a radical new toolkit, shifting from analytic design to learning codes directly from data. By parameterizing the encoder and decoder as DNNs and optimizing them end-to-end, DL-based feedback codes can learn to exploit the feedback link by capturing complex, non-linear dependencies across transmission rounds that are difficult to engineer analytically.

Pioneering work in this domain, such as DeepCode \cite{DeepCode}, utilized recurrent neural networks to leverage bit-by-bit passive feedback, where the receiver feeds back its noise-corrupted received symbols. This was advanced by AttentionCode \cite{AttentionCode}, which replaced RNNs with attention-based DNN to model longer-range dependencies, achieving lower block error rates. The current state-of-the-art in this lineage is represented by Generalized Block Attention Feedback (GBAF) codes \cite{GBAF}, which employ the attention-based DNN for efficient group-wise encoding and decoding, setting new performance benchmarks under passive feedback. A lightweight neural coding scheme was proposed in \cite{ankireddy2025lightcode}, leveraging the same group-wise decoding approach.

This progress has spurred several meaningful extensions. Research has shown that adapting the GBAF architecture for active feedback \cite{active}, where the receiver processes its observation before generating a feedback signal, can yield further gains when the feedback channel is noisy. Other innovations include variable-length feedback strategies \cite{DeepVLF}, which aim to improve spectral efficiency by dynamically adjusting the communication length based on channel conditions. Collectively, these works have demonstrated that DNNs can synthesize feedback utilization strategies that surpass conventional HARQ in controlled, simulated settings.

\subsection{Contributions}
Existing feedback codes, while promising in theory and simulations, remain largely ``point solutions'' designed for idealized conditions. 
Their development often overlooks three critical and interrelated practical challenges:
\begin{itemize}[leftmargin=0.5cm]
\item \textit{Fixed-channel training:} Existing schemes are designed or trained at a fixed channel condition. In practice, the received signal-to-noise ratio (SNR) fluctuates continuously, causing severe performance degradation for models overfitted to a static operating point.
\item \textit{Ignored latency and synchrony:} Existing designs typically assume instantaneous encoding/decoding and zero-delay feedback, neglecting the computational latency of DNN inference, protocol processing, and wireless scheduling. These cumulative delays can easily violate stringent air-interface timing constraints in practice.
\item \textit{High encoder complexity:} Many approaches employ encoders with large, computationally intensive DNNs, which are ill-suited for battery-limited, resource-constrained end-user devices, undermining deployment feasibility.
\end{itemize}

To bridge this substantial ``simulation-to-reality'' gap, this paper introduces Rich-ARQ, a unified framework designed to operate efficiently and reliably under the real-world constraints. Our work makes the following key contributions:
\begin{enumerate}[leftmargin=0.5cm]
\item To tackle the challenge of delayed feedback, we introduce a fundamental re-conceptualization of feedback coding, moving from synchronous designs, where the encoder stalls waiting for the latest feedback, to a novel asynchronous paradigm. We put forth a new asynchronous feedback code (AFC), which can generate new parity symbols using historical feedback, without being blocked by delayed or missing feedback. This architectural shift creates a robust, overlapping pipeline that drastically reduces latency compared to existing feedback codes.
\item To combat performance collapse under time-varying channels, we develop an SNR-conditioned curriculum learning method with Langevin perturbations. This training strategy mimics the stochastic fluctuations of real channel SNR, enabling AFC to learn adaptive coding and feedback policies across a wide range of dynamic channel conditions.
\item To address the high-complexity barrier at resource-constrained devices, we engineer a lightweight AFC encoder through model pruning and sparse computation. Rich-ARQ adopts a pragmatic asymmetric design that aligns with the resource asymmetry of the star topology: a lightweight encoder resides at the user equipments (UEs), while a powerful decoder operates at the resource-rich access point (AP).
\item We build, to our knowledge, the first operational feedback channel coding prototype, a full-stack software-defined radio (SDR) system with a standard-compliant 4G/5G physical layer. Its key innovation is a non-blocking, deadline-aware execution architecture that decouples DNN inference from strict physical-layer timing, resolving the core system challenge of integrating variable-latency AI with real-time radio.
\end{enumerate}

Through comprehensive over-the-air experiments, we demonstrate that Rich-ARQ delivers superior performance over both conventional HARQ and prior DL-based feedback codes in terms of spectral efficiency, coverage, and reliability. Specific gains include an $8.8$-$9.5$ dB lower SNR than Turbo-HARQ and Polar-HARQ, respectively, required to achieve a target PER of $10^{-4}$. This translates to a $1.38\times$ and $1.70\times$ increase in maximum communication distance relative to the two HARQ baselines.
Moreover, Rich-ARQ maintains robust performance across varying channel conditions where prior DL-based codes degrade sharply, and achieves a $43.4\%$ reduction in end-to-end latency compared with state-of-the-art DL-based feedback code. This work moves feedback coding from theory to practice, offering a viable new paradigm for next-generation ultra-reliable wireless communication systems.

\section{System Model}\label{sec:Multi_Access}
\begin{figure}[t]
\centering
\includegraphics[width=0.45\textwidth]{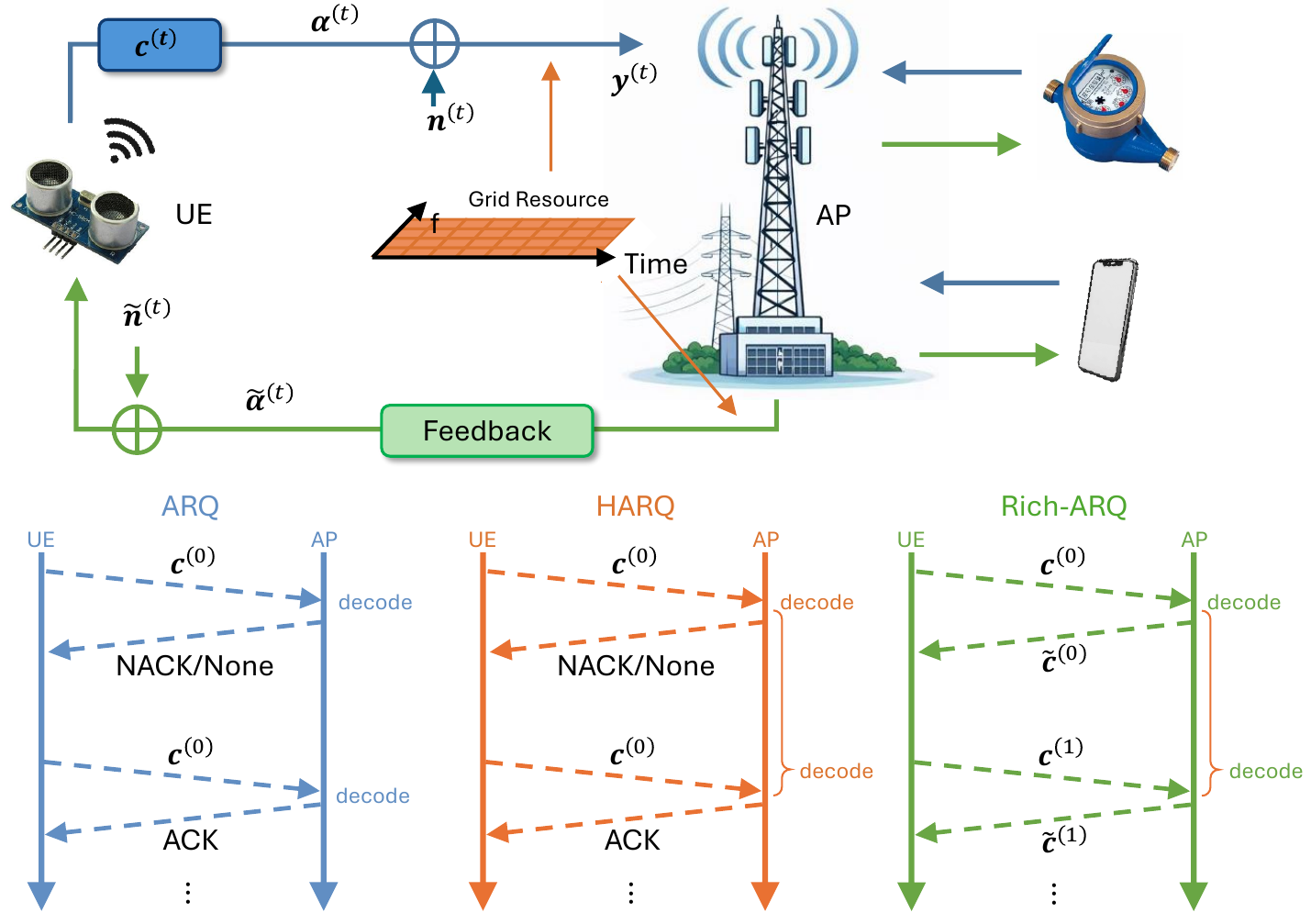}
\caption{The prevailing star-topology architecture in contemporary wireless systems, and a comparison among ARQ, HARQ, and Rich-ARQ.}
\label{fig_multi_user}
\end{figure}

We consider the prevailing architectural paradigm: the {star topology}, in contemporary wireless systems such as 4G/5G, Wi-Fi, and IoT networks. As illustrated in Fig.~\ref{fig_multi_user}, a central AP or base station communicates with multiple UEs in its coverage area. This topology exhibits a fundamental {asymmetry} between the two ends \cite{DeepIoT}: the AP is typically grid-powered, equipped with multiple antennas, and possesses substantial computational resources; in contrast, UEs are often battery-driven, equipped with a single antenna, and operate under energy constraints. Consequently, the downlink channel quality is typically superior to that of the uplink \cite{kanj2020tutorial}.

In such systems, the time-frequency radio resources are partitioned into structured units called {Physical Resource Blocks (PRBs)}. The AP centrally schedules these PRBs among the UEs, allocating specific slots and subcarriers for each uplink (UE $\to$ AP) or downlink (AP $\to$ UE) transmission.

\subsection{Automatic Repeat Request (ARQ)}
Without loss of generality, we focus on a single UE transmitting a data packet to the AP. To transmit a block of $K$ information bits, $\bm{b}\in\{0,1\}^K$, the UE first performs channel coding (e.g., Turbo or Polar coding) and modulation, producing a sequence of $M$ coded symbols denoted by $\bm{c}^{(0)}$. These symbols are then mapped onto subcarriers in the allocated PRBs, and an Orthogonal Frequency-Division Multiplexing (OFDM) waveform is generated for over-the-air transmission.

The signal received at the AP on the $i$-th subcarrier can be written as
\begin{equation}
    \label{eq_cny}
    {y}_i^{(0)}= \alpha^{(0)}_i \cdot c_i^{(0)}+n_{i}^{(0)},
\end{equation}
where $\alpha^{(0)}_i$ and $n_{i}^{(0)}$ are the frequency-domain response and noise of the uplink channel, respectively.
Upon receiving the signal, the AP attempts to decode the packet. In standardized feedback protocols, encompassing both conventional ARQ and more advanced HARQ, the information returned to the UE is minimalistic: if decoding succeeds, the AP sends a one-bit ACK; in case of failure, the feedback is typically implicit, e.g., the absence of an ACK triggers a timeout, or explicit, e.g., a one-bit NACK is sent, as in many HARQ implementations. Crucially, regardless of the specific mechanism, the informational content of the feedback remains binary: a coarse verdict of success or failure.

In the event of a NACK (or absence of ACK), the UE initiates a retransmission, as shown in Fig.~\ref{fig_multi_user}.
In conventional ARQ, the retransmitted packet $\bm{c}^{(t)}$ (where the superscript $t$ indexes the transmission attempt) is typically an identical replica of the original. 
In HARQ schemes, $\bm{c}^{(t)}$ can be either a replica (chase combining) or a new set of parity bits (incremental redundancy), which the AP combines with prior receptions to enhance decoding \cite{frenger2001performance}. 
While this framework forms the backbone of reliable wireless communication, it reduces the role of feedback to a one-bit control signal, discarding all nuanced information about channel state, decoder confidence, or the nature of the residual errors.

\subsection{Rich-ARQ}
The above observation motivates the core concept of our work: extending the feedback from a single bit to a multi-dimensional, coded packet. In the proposed Rich-ARQ framework, the feedback sent in the $t$-th round, denoted by $\bm{\widetilde{c}}^{(t)}$, $t=0,1,2,...$, is generated at the AP by processing the received signal $\bm{y}^{(t)}$ to extract a compact yet informative feature vector. This vector $\bm{\widetilde{c}}^{(t)}$ is then transmitted back to the UE via the downlink. 

The signal received at the UE on the $i$-th subcarrier can be written as
\begin{equation}
    \widetilde{y}_i^{(t)}= \widetilde{\alpha}^{(t)}_i \cdot \widetilde{c}_i^{(t)}+\widetilde{n}_{i}^{(t)},
\end{equation}
where $\widetilde{\alpha}^{(t)}_i$ and $\widetilde{n}_{i}^{(t)}$ are the frequency-domain response and noise of the feedback channel, respectively.
Upon reception, the UE's encoder utilizes this rich feedback $\bm{\widetilde{y}}^{(t)}$, together with the original message bits $\bm{b}$ and previously transmitted symbols $\bm{c}^{(\tau)}$, $\tau=0,1,2...,t$, to produce the next coded transmission $\bm{c}^{(t+1)}$. This establishes a closed-loop, collaborative coding cycle where the feedback actively informs the forward encoder, enabling it to adaptively refine subsequent transmissions based on the decoder's observations.

Translating the Rich-ARQ framework into a practical system hinges on the design of an efficient feedback channel code. However, existing feedback codes are predominantly developed and evaluated under highly idealized conditions, relying on assumptions that are often invalid in real-world wireless deployments.

\begin{figure}[t]
\centering
\includegraphics[width=0.4\textwidth]{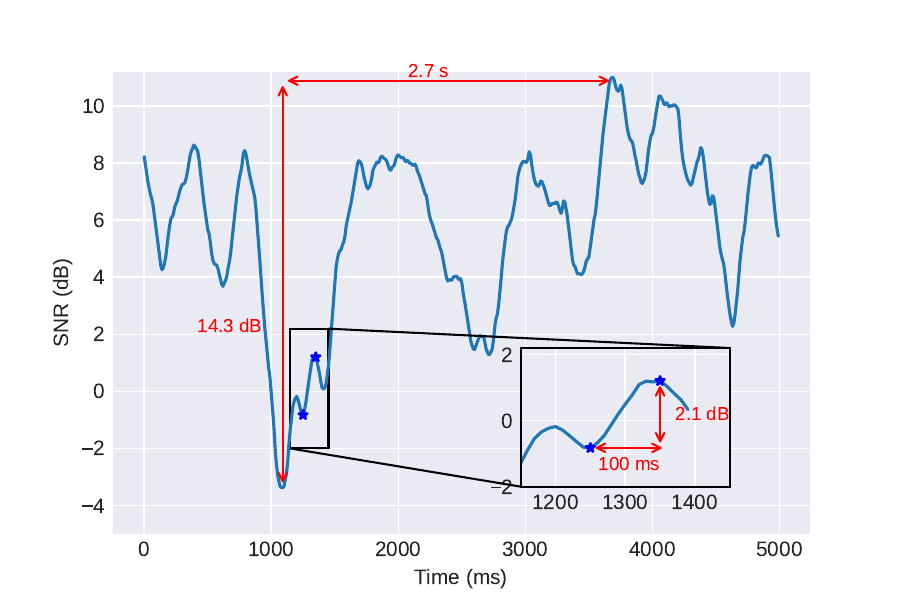}
\caption{The measured SNR variation in an indoor laboratory environment.}
\label{fig_snr}
\end{figure}

A primary limitation is the assumption of a fixed SNR \cite{SK1,DeepCode}. Existing feedback codes (both classical and DL-based) are typically designed and optimized for a single, predetermined SNR. In practice, SNR varies continuously due to mobility, interference, and environmental changes. Fig.~\ref{fig_snr} illustrates the measured SNR variation in an indoor laboratory environment (the underlying OFDM transceiver is detailed later in Section~\ref{sec:system_impl}). As shown, the received SNR fluctuates substantially: over a $100$ ms interval, the variation can reach $2.1$ dB; over $2.7$ s, it can exceed $14.3$ dB. Codes optimized for a single, predetermined SNR typically suffer significant performance degradation when subjected to such realistic, time-varying conditions.

A second critical issue is the ignorance of coding latency \cite{Shannon,GBAF}. Existing code designs frequently assume instantaneous encoding and decoding, neglecting the computational time required in practice. When implemented on real hardware, especially for DL-based schemes, the encoding and decoding steps introduce non-negligible delays. If operated with existing feedback codes, these accumulated computation delays over multiple transmission rounds can easily violate strict air-interface timing deadlines. Furthermore, in multi-user systems with limited spectral resources, scheduling and transmitting the richer feedback packets themselves introduce additional communication delay, which must be accounted for in the overall system design.

Furthermore, high computational complexity poses a practical barrier \cite{GBAF,active}. In particular, the class of DL-based feedback codes often involve DNNs with considerable parameter counts and operations, making them ill-suited for resource-constrained UEs (e.g., IoT sensors or smartphones) where energy and memory are limited. Prior schemes have largely overlooked the need to reduce the encoder's computational burden for on-device deployment.

These obstacles render existing feedback codes impractical for realizing Rich-ARQ in real wireless systems. To bridge this gap, we introduce a new class of practical feedback code in the next section to achieve the vision of Rich-ARQ.

\section{Asynchronous Feedback Code: Enabling Practical Rich-ARQ}\label{sec:AFC}
Translating the Rich-ARQ vision into practice requires overcoming the limitations outlined in Section~\ref{sec:Multi_Access}. In this section, we present the solution: the Asynchronous Feedback Code (AFC), a novel feedback coding framework designed around three core principles: i) an asynchronous design that eliminates encoder stalls, ii) native adaptation to varying channel SNR, and iii) a lightweight encoder suitable for resource-constrained devices. Together, these innovations make the Rich-ARQ paradigm practical for real-world deployment.

\begin{figure*}[t]
    \centering
    \includegraphics[width=0.7\textwidth]{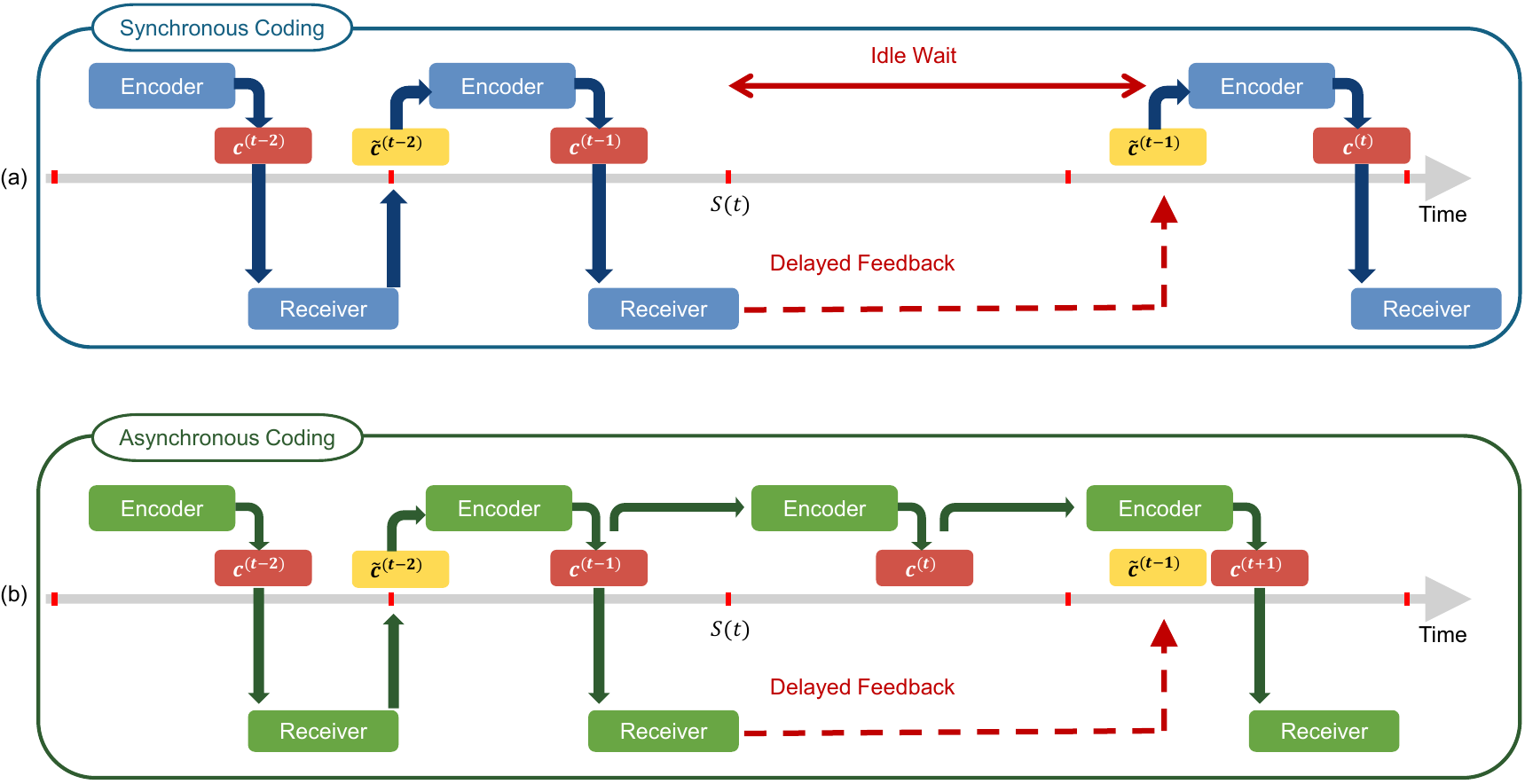}
    \caption{Time line comparison of feedback coding: (a) Synchronous coding timeline, where each transmission round must await feedback from the previous round before proceeding. (b) Asynchronous coding timeline, where the encoder generates the next codeword $\bm{c}^{(t)}$ despite delayed feedback $\widetilde{\bm{c}}^{(t-1)}$, eliminating idle waiting and reducing end-to-end latency.}
    \label{fig_time_line}
\end{figure*}

\subsection{Asynchronous Coding}

In real-world communication networks, feedback is never instantaneous. Processing delays at the receiver, queuing within network stacks, and contention for downlink transmission slots collectively ensure that the acknowledgment for a transmitted packet arrives at the sender with variable and often unpredictable latency. This fundamental reality of feedback delay exposes a critical, yet overlooked, assumption in prior feedback code designs: the strict temporal alignment, or synchrony, between forward transmissions and their corresponding feedback.

This observation leads us to propose a foundational re-categorization of feedback codes based on their temporal dependency. We classify all existing schemes as \textit{synchronous} feedback codes. In these schemes, the encoder's operation is gated by feedback arrival. To generate the codeword for transmission round $t+1$, the encoder must first receive and process the feedback packet specifically generated in response to its $t$-th transmission. This creates a lock-step, sequential pipeline: transmit($t$) $\to$ wait $\to$ receive feedback($t$) $\to$ compute $\to$ transmit($t+1$). Any delay in the feedback loop directly translates to idle waiting at the encoder, stalling the entire process and accumulating into significant overall latency, as illustrated in the upper timeline of Fig.~\ref{fig_time_line}.

The core innovation of AFC is to shatter this synchronous paradigm. AFC is, by design, an \textit{asynchronous} feedback code. Its encoder is liberated from the need to wait for the most recent feedback. Instead, it operates on a continuous, opportunistic basis, generating new parity symbols using any and all information that has already arrived. Formally, when constructing the codeword $\bm{c}^{(t)}$ for the $t$-th transmission, the AFC encoder utilizes:
\begin{itemize}
    \item The original message bits $\bm{b}$;
    \item All previously transmitted codewords $\{\bm{c}^{(\tau)}\}_{\tau=0}^{t-1}$;
    \item The set of all feedback signals ${\widetilde{\bm{y}}^{(\tau)}}$ that have been received up to that moment.
\end{itemize}

Crucially, the index of the latest available feedback, denoted $t'$, may be less than $t-1$ (i.e., feedback for transmission $t-1$ might still be in transit). The encoder proceeds regardless. It simply treats the missing feedback vectors $\widetilde{\bm{y}}^{(t'+1)}, ..., \widetilde{\bm{y}}^{(t-1)}$ as ``not yet received'' and excludes them from the current encoding step. Its internal state is an accumulation of the entire history of interaction, not just the last exchange.

The power of this approach is demonstrated in Fig.~\ref{fig_time_line}. Consider the moment when feedback $\widetilde{\bm{c}}^{(t-1)}$ is delayed. A synchronous encoder is forced to idle at point $S(t)$, its pipeline frozen. In contrast, the AFC encoder has already consumed the earlier, on-time feedback $\widetilde{\bm{c}}^{(t-2)}$ to produce $\bm{c}^{(t)}$, without waiting for $\widetilde{\bm{c}}^{(t-1)}$. By decoupling the forward transmission timeline from the feedback reception timeline, AFC transforms the coding process from a fragile chain of sequential dependencies into a robust, overlapping pipeline. This architectural shift inherently tolerates variable feedback delay, efficiently exploits all received feedback information, and, most critically, achieves a dramatic reduction in end-to-end latency, making the Rich-ARQ paradigm truly practical for time-sensitive wireless systems.

\subsection{SNR-Robust Adaptive Coding}
The second pillar of AFC addresses a fundamental, yet often overlooked, reality of wireless channels: their inherent non-stationarity. As captured in Fig.~\ref{fig_snr}, the channel SNR is not a fixed parameter but a continuously fluctuating variable. Over the multi-round interaction of a Rich-ARQ session between the UE and AP, the SNR can traverse a wide dynamic range. A practical feedback code must, therefore, be inherently adaptive, capable of dynamically adjusting its coding strategy in response to real-time channel conditions to maintain optimal performance. This requirement poses a profound challenge, particularly for the transmitter: \textit{over successive rounds, how can the encoder discern whether a decoding failure stems from a severe channel degradation (a sustained drop in $\alpha_i$) or merely an unfortunate noise realization (a large but transient $n_i$)}? Relying solely on the binary success/failure signal or even multi-dimensional feedback about the received symbols leaves this ambiguity unresolved, crippling the encoder's ability to make informed decisions.

To resolve this ambiguity and empower the encoder with channel state awareness, we introduce a critical yet low-overhead addition to the rich feedback: the AP estimates and feeds back the instantaneous channel SNR, $\gamma^{(t)}$, for each transmission round. This scalar metric serves as a direct, high-level summary of the channel condition during that specific transmission. At the UE, this SNR value is not used raw; instead, it is processed by a compact, shared Multilayer Perceptron (MLP) that transforms it into a high-dimensional channel-state embedding, $\gamma^{(t)}_{\text{emb}}$. This embedding is then concatenated with the other inputs (the message bits, history of transmitted symbols, and historical feedback) at the input stage of the neural encoder and decoder, as illustrated in Fig.~\ref{fig_light}. This mechanism grants the DNN an explicit, contextual signal about the propagation environment, enabling it to learn distinct coding and feedback-generation policies for high-SNR (aggressive, spectral-efficient) versus low-SNR (robust, redundancy-focused) regimes.

However, merely providing SNR as an input during inference is insufficient. The DNN must be trained to generalize robustly across the entire continuum of possible SNR values it might encounter, from excellent to deeply faded conditions. Training on a fixed SNR leads to a model specialized for that point, which catastrophically fails under variation. Training on a broad, uniform distribution of SNRs often results in a model that performs inadequately across the board, failing to master the subtleties of any specific regime.

The innovation of AFC lies in a new SNR-conditioned curriculum learning with Langevin perturbations \cite{bengio2009curriculum,welling2011bayesian}. We frame the problem through the lens of stochastic dynamics: the observed SNR trace in Fig.~\ref{fig_snr} is not merely a curve but a realization of a stochastic process. The temporal fluctuations measured in practice share a key statistical characteristic with the principle of Langevin dynamics: the presence of a deterministic drift (e.g., due to path loss or slow fading) superposed with random perturbations (due to multipath, interference, and measurement noise).

We therefore design our training protocol to mimic and exploit this structure. We define two anchor SNR distributions: a benign origin distribution, $P_{\text{orig}}(\gamma)$, representing favorable conditions (e.g., high mean), and a challenging target distribution, $P_{\text{targ}}(\gamma)$, representing harsh conditions (e.g., low mean). Instead of abruptly switching between them, we guide the model through a smooth, curriculum-driven transition. During training, the SNR for each batch, $\gamma_{\text{train}}$, is generated as
\begin{equation}
\label{eq_snr_mix}
\gamma_{\text{train}} = \gamma_{\text{mix}} + \Gamma_{\text{pert}},
\end{equation}
where the base SNR $\gamma_{\text{mix}}$ is drawn from a time-varying mixture of the two anchor distributions:
\begin{equation}
\gamma_{\text{mix}} \sim \alpha(k) P_{\text{orig}}(\gamma) + \left[1-\alpha(k)\right] P_{\text{targ}}(\gamma).
\end{equation}
Here, $\alpha(k) \in [0,1]$ is a curriculum coefficient that decays from nearly 1 to 0 as the training progress $k$ increases, gradually shifting the focus from easy to hard examples.

The additive term $\Gamma_{\text{pert}} \sim \mathcal{N}(0, \sigma_p^2)$ in \eqref{eq_snr_mix} is an independent Langevin-style random perturbation. This serves a dual purpose: 1) Practical match: It directly emulates the random fluctuations observed in real-world SNR measurements (Fig.~\ref{fig_snr}), ensuring the model is trained on realistically noisy channel state information; 2) Theoretical benefit: It acts as a regularizer during the curriculum transition. As the mean of the mixture distribution drifts towards lower SNRs, these perturbations ensure that the model's parameter updates are exposed to a local neighborhood of SNR values around the current mean. This prevents catastrophic forgetting of higher-SNR strategies and promotes the learning of a smooth, generalized coding strategy that is robust to small, stochastic SNR variations around any operating point. The variance $\sigma_p^2$ controls the exploration radius in the SNR space, linking the training noise directly to the expected volatility of the real channel.

This SNR-conditioned curriculum learning approach moves beyond simply ``training with variable SNR''. It is a principled methodology that co-designs the learning process with the statistical physics of the channel, ensuring that the resulting AFC model does not just see many SNRs, but learns to navigate the SNR landscape intelligently and robustly, as required for practical deployment in fluctuating environments.

\subsection{Lightweight Coding}

\begin{figure}[t]
\centering
\includegraphics[width=0.3\textwidth]{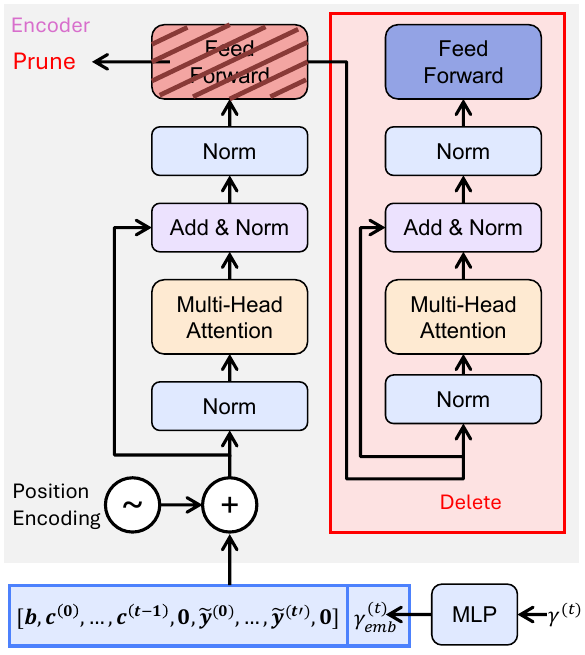}
\caption{Lightweight AFC encoder architecture implemented at the UE to meet hardware constraints.}
\label{fig_light}
\end{figure}

The practical deployment of any neural coding scheme hinges on its computational footprint, particularly at the resource-constrained transmitter (UE). The final component of the AFC design, therefore, focuses on creating a lightweight encoder that retains the core functionalities of asynchrony and SNR-adaptation while being executable on mobile hardware with limited processing power, memory, and energy.

The AFC encoder employs a neural architecture based on our prior work in DL-based feedback codes, which utilizes attention mechanisms to effectively manage sequential coding decisions. As shown in Fig.~\ref{fig_light}, the encoder at the UE integrates diverse information into a structured input matrix at each step $t$:
\begin{equation}
\label{eq_enc_input}
\bm{Q}^{(t)}_{\text{enc}} = [\bm{b}, \bm{c}^{(0)}, \ldots, \bm{c}^{(t-1)}, \bm{0}, \widetilde{\bm{y}}^{(0)}, \ldots, \widetilde{\bm{y}}^{(t')}, \bm{0}, \gamma^{(t)}_{\text{emb}}],
\end{equation}
where $\gamma^{(t)}_{\text{emb}}$ is the embedding of the current SNR. This matrix is processed by a stack of neural layers. Each layer typically performs a sequence of operations including layer normalization, a single-head self-attention mechanism, and a feed-forward transformation, ultimately generating the next parity symbols $\bm{c}^{(t)}$.

To ensure this encoder is viable for on-device execution, we deliberately streamline its architecture through two primary methods:
\begin{itemize}[leftmargin=0.5cm]
    \item Model pruning \cite{reed1993pruning}: We reduce the encoder's complexity by employing a shallower architecture, using fewer neural layers and smaller internal feature dimensions compared to the decoder. This directly addresses constraints on memory and computation at the UE.
    \item Sparse feed-forward computation: The original feed-forward mechanism has considerable complexity, which becomes prohibitive for multiple transmission rounds. We implement a sparse feed-forward pattern that restricts interactions to the most relevant dependencies, for instance, primarily linking recent feedback to current parity generation. This significantly lowers the computational load, as denoted by the slashed blocks in Fig. \ref{fig_light}.
\end{itemize}

These design choices are guided by a clear operational principle: the UE encoder's primary task is to efficiently generate the next parity symbols by synthesizing the available context (message, transmission history, and received feedback), not to perform the more complex inference required for decoding. The streamlined architecture is sufficient to maintain the essential stateful behavior needed for asynchronous operation and SNR-conditioned adaptation.

In contrast, the decoder at the resource-rich AP retains a full-capacity, multi-layer architecture. Its roles are more computationally demanding: it must not only generate the rich feedback vector $\widetilde{\bm{c}}^{(t)}$ after each reception but also ultimately decode the original message $\bm{b}$ after $T$ rounds. For feedback generation, the decoder's input includes the latest received symbol and the SNR. For final decoding, it integrates the sequence of all received symbols $\bm{y}^{(0)}, \ldots, \bm{y}^{(T-1)}$ and the SNR embedding. This requires a more powerful network capable of synthesizing information across the entire transmission history and performing complex inference.

We emphasize that this asymmetric complexity design of AFC is a deliberate and pragmatic choice that mirrors the inherent resource asymmetry of the star-network topology. The UE runs a lightweight encoder whose primary role is efficient, context-aware parity generation. The AP hosts a more powerful decoder responsible for the computationally intensive tasks of feedback synthesis and final decoding. This division of labor ensures that the collaborative intelligence of Rich-ARQ is achieved without imposing a prohibitive computational burden on the mobile device, making the AFC framework both high-performing and practical for real-world deployment.

\section{The Rich-ARQ Prototype}
\label{sec:system_impl}

\subsection{Hardware Platform and Physical Layer} 
\begin{figure*}[t]
\centering
\includegraphics[width=0.7\textwidth]{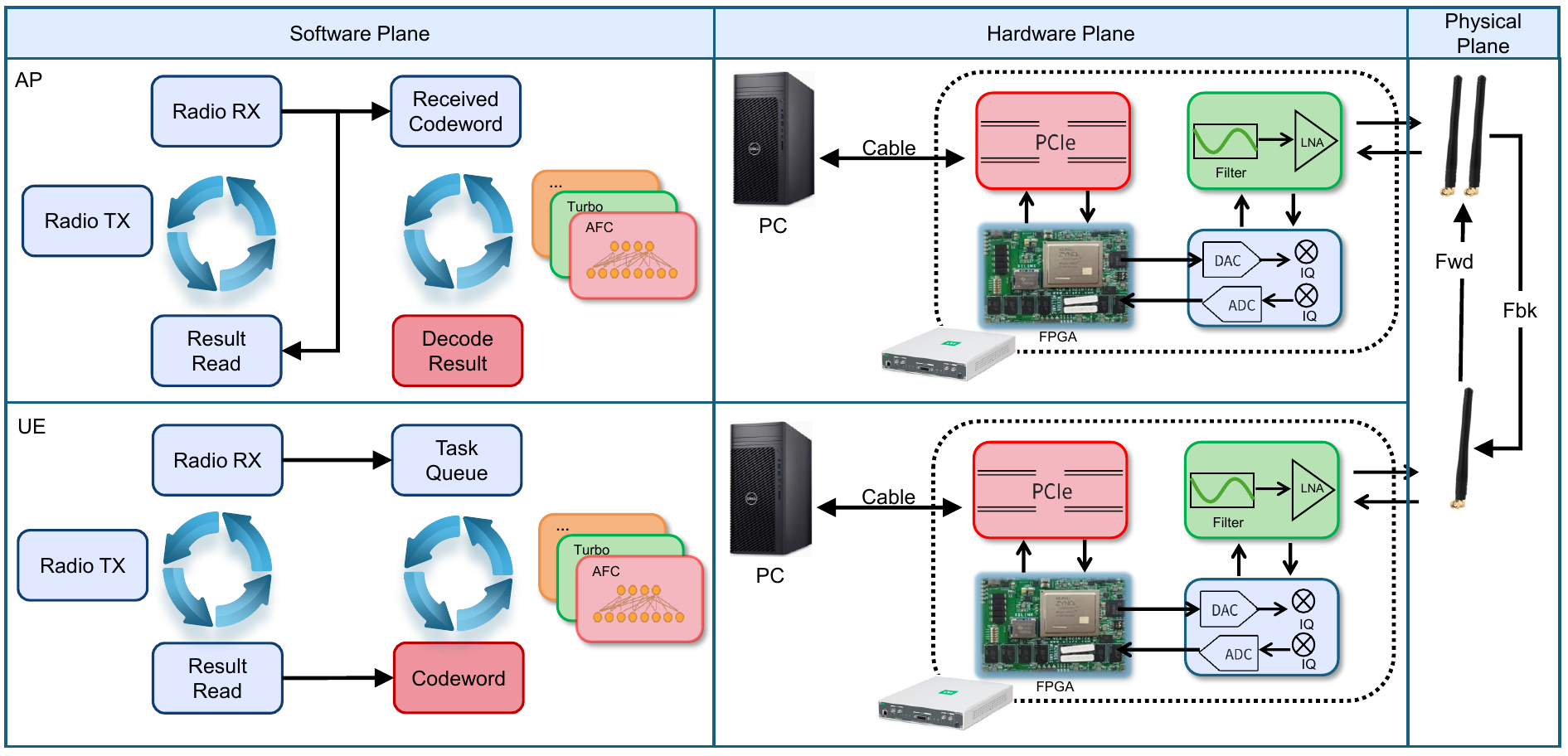}
\caption{Hardware and software architecture of the Rich-ARQ prototype system. The diagram depicts both the AP and a UE, each comprising a host computer running the layered protocol stack and the AFC encoder and decoder, connected via PCIe and Ethernet to a USRP X310 software-defined radio frontend. The software plane illustrates the real-time, multi-threaded execution architecture that decouples neural network inference from time-critical physical layer processing.}
\label{fig_prototype}
\end{figure*}

Having introduced the algorithmic principles of the AFC, we now present its materialization into a functional, real-world system. This section details the implementation of our Rich-ARQ prototype, a complete software-defined radio (SDR) platform designed to validate the core proposition: that a high-dimensional, neural-coded feedback loop can be reliably operated within the stringent timing and structural confines of contemporary wireless standards like 4G LTE and 5G NR. The development of this prototype necessitated solving several critical engineering challenges, chief among them the reconciliation of variable-latency neural network inference with microsecond-precise radio deadlines. Our prototype, as depicted in Fig.~\ref{fig_prototype}, demonstrates that Rich-ARQ is not merely a simulation model but a viable communication protocol ready for over-the-air experimentation.

Our prototype instantiates the asymmetric star topology central to modern networks. It consists of an AP and several UEs. As shown in Fig. \ref{fig_prototype}, both the AP and UE nodes share an identical hardware composition for flexibility: a general-purpose server (Intel i9-14900K, 32GB RAM) is connected via a 10 Gigabit Ethernet link to a USRP X310 SDR equipped with a UBX-160 daughterboard. This setup provides the necessary computational power for the DNNs and the RF performance for wideband operation. A shared 10 MHz reference clock synchronizes the local oscillators of both USRPs, ensuring carrier frequency coherence, a prerequisite for stable OFDM demodulation. The software stack, including the physical layer processing and the AFC neural models, is implemented in C++.

\begin{figure*}[t]
\centering
\includegraphics[width=0.7\textwidth]{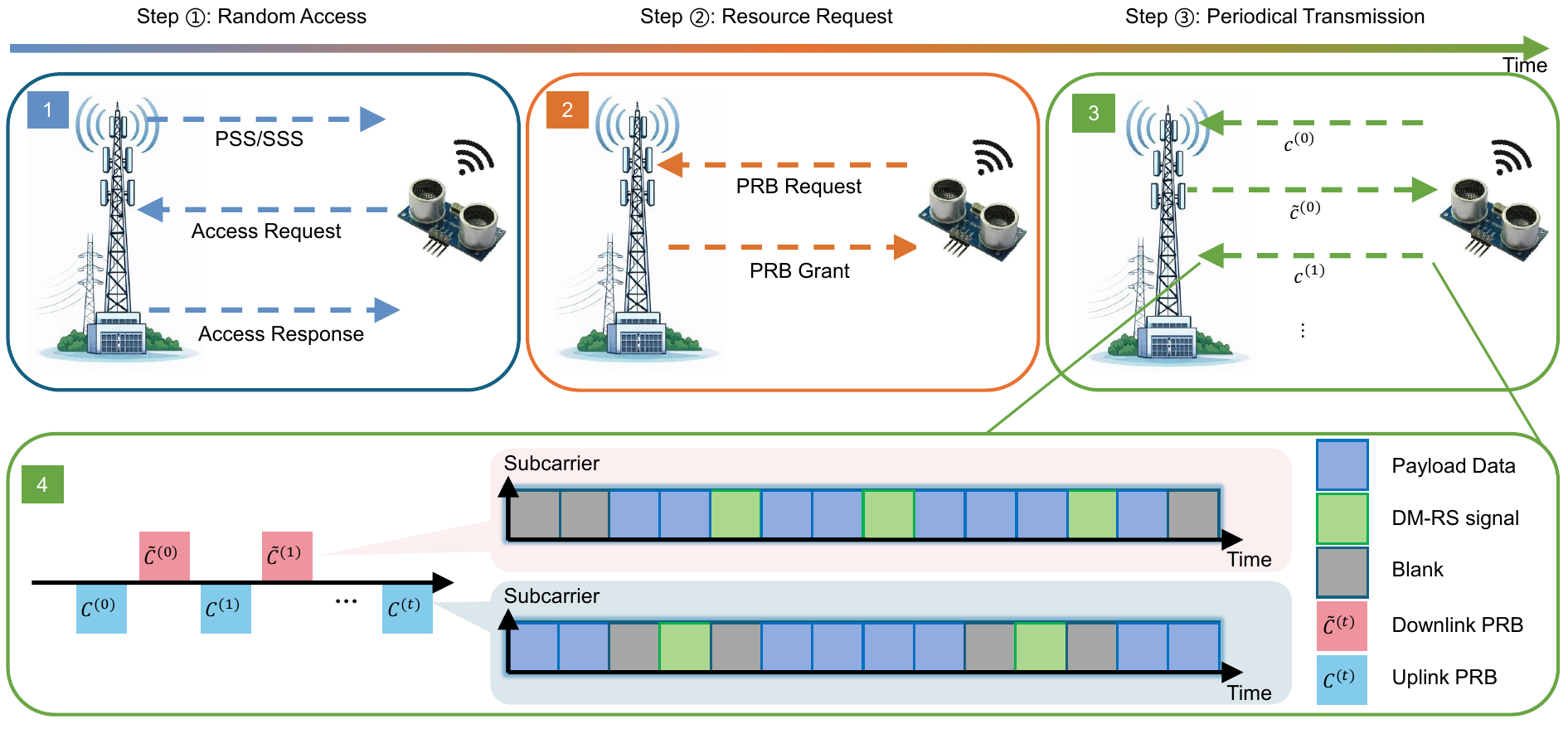}
\caption{Protocol timeline and physical layer structure of the Rich-ARQ system: 1) Link establishment via downlink synchronization and uplink random access; 2) Scheduling phase where the AP grants periodic uplink resources; 3) Multi-round Rich-ARQ interactive session, where high-dimensional coded feedback replaces conventional 1-bit ACK/NACK; 4) Detailed mapping of DM-RS and REs within a PRB, illustrating the transmission of AFC parity symbols (uplink) and rich feedback symbols (downlink).}
\label{fig_access_and_grid}
\end{figure*}

To ensure our work is grounded in practical deployment scenarios, the prototype's physical layer is designed to be fully compatible with LTE and NR frameworks. We adopt the standard PRB as the fundamental time-frequency unit. The resource mapping within a PRB is illustrated in Fig. \ref{fig_access_and_grid}. Each PRB spans 1 ms (one subframe comprising 14 OFDM symbols) in time and 180 kHz (12 subcarriers at 15 kHz spacing) in frequency \cite{kanj2020tutorial}. This grid is populated with:
\begin{itemize}[leftmargin=0.5cm]
\item \textit{Demodulation Reference Signals (DM-RS):} These are known pilot symbols inserted at specific positions, used by the receiver for channel estimation and equalization of the data symbols.
\item \textit{Data Resource Elements (REs):} These REs carry the payload. In the uplink, they transmit the AFC-generated parity symbols $\bm{c}^{(t)}$. In the downlink, they carry the rich feedback vectors $\widetilde{\bm{c}}^{(t)}$.
\end{itemize}
This structure is identical to that used for standard data transmission in 5G NR, meaning a Rich-ARQ session occupies the network's resources in a perfectly standard-compliant manner.

\subsection{Protocol Procedure}

The interaction between the UE and AP follows a modified yet backward-compatible protocol sequence, contrasting sharply with the minimalism of conventional ARQ. The procedure, outlined in Fig. \ref{fig_access_and_grid}, consists of three phases:

\subsubsection{Standard link establishment} This phase is unchanged from 5G NR \cite{rinaldi20215g}. The AP first broadcasts synchronization signals (PSS/SSS). The UE detects these to achieve downlink time and frequency synchronization. Subsequently, the UE initiates a random-access procedure by transmitting a preamble. The AP measures the preamble's arrival time and responds with a ``Timing Advance'' command, which the UE uses to align its future uplink transmissions, compensating for propagation delay.

\subsubsection{Resource grant for Rich-ARQ} The UE sends a scheduling request to the AP. The AP's scheduler, aware that this UE will engage in a multi-round Rich-ARQ session, responds with a grant allocating a series of periodic PRBs for uplink transmission. This scheduling is identical to granting resources for a large data packet in a standard system.

\subsubsection{The Rich-ARQ interactive session} This is where the paradigm shift occurs. Instead of transmitting a large, static codeword, the UE begins the AFC process. It transmits the initial parity block $\bm{c}^{(0)}$ in the first granted PRB.
\begin{itemize}[leftmargin=0.5cm]
\item {Traditional ARQ/NACK:} The AP would simply attempt to decode. If successful, it sends a 1-bit ACK. If not, it either sends a 1-bit NACK or remains silent (triggering a timeout). The UE's only recourse is a blind retransmission.
\item {Rich-ARQ:} Upon receiving $\bm{y}^{(0)}$, the AP's AFC decoder does \textit{not} make a final binary decision. Instead, it processes $\bm{y}^{(0)}$ to generate a multi-dimensional feedback vector $\widetilde{\bm{c}}^{(0)}$. This vector is a coded representation of the decoder's state (i.e., its confidence, estimated channel distortions, or residual uncertainty). $\widetilde{\bm{c}}^{(0)}$ is then transmitted back to the UE in the downlink. The UE's AFC encoder uses this rich feedback, along with the message bits and the previous transmission, to generate a refined, adaptive parity block $\bm{c}^{(1)}$ for the next granted PRB. This iterative, information-rich interaction continues for a predefined number of rounds.
\end{itemize}

Thus, Rich-ARQ co-opts the standard resource grant mechanism but replaces the simplistic ACK/NACK-and-retransmit logic with a coordinated, multi-round, stateful coding interaction. The physical layer framing remains standard; the intelligence lies in the content of the data and feedback fields.

\subsection{Real-Time Execution Architecture}
\label{subsec:exe_archi}

A paramount challenge in deploying neural network-based codes on real radios is their unpredictable and often substantial inference latency. A radio physical layer (PHY) operates on a strict timeline; missing a transmission deadline results in a corrupted frame or dead air. A naive design that blocks the PHY thread waiting for the neural encoder's output is therefore infeasible.

Our key systems innovation is a non-blocking, deadline-aware software architecture, depicted in the software plane of Fig. \ref{fig_prototype}. This architecture physically decouples the neural network processing from the real-time signal processing pipeline through a multi-threaded design with protective safeguards.

\subsubsection{Encoder at the UE} 
The real-time PHY thread handles all time-critical sample processing, including OFDM modulation, cyclic prefix insertion, and managing the USRP's transmit buffer. To accommodate the variable execution time of the neural encoder, the system employs a non-blocking, deadline-driven workflow. When the next transmission opportunity (e.g., for $\bm{c}^{(t+1)}$) is imminent (typically $2$-$3$ ms in advance), the PHY thread asynchronously dispatches a codeword request to a dedicated ``encoder worker thread'', which hosts the AFC neural network. The PHY thread then continues its other tasks without waiting.

At the precise moment when the transmit buffer must be finalized, the PHY thread checks a shared memory location for the newly generated codeword.
\begin{itemize}[leftmargin=0.5cm]
\item If the codeword $\bm{c}^{(t+1)}$ is ready, it is immediately committed for transmission.
\item If the result is not yet available, the transmission slot is left empty (i.e., no symbols are sent). This safeguard is distinct from the algorithmic concept of asynchronous coding described in Section~\ref{sec:AFC}; it is a system-level robustness measure. Since the prototype runs on a general-purpose operating system where thread scheduling and neural network inference times can be unpredictable, this mechanism ensures that the PHY layer never misses its strict timing deadlines, even if the encoder experiences occasional delay.
\end{itemize}

\subsubsection{Decoder/Feedback at the AP} A similar decoupling is implemented. Upon receiving a frame, the PHY thread immediately forwards the symbols to a \textit{Feedback Generation Thread}. This thread performs low-latency operations (channel estimation, SNR calculation) and runs a streamlined portion of the AFC decoder to rapidly produce the rich feedback vector $\widetilde{\bm{c}}^{(t)}$, which is then scheduled for downlink transmission with minimal delay. Concurrently, the received symbols are passed to a more powerful \textit{Main Decoder Thread} which runs the full AFC decoder asynchronously, accumulating the history $\bm{y}^{(0:T-1)}$ to eventually decode the original message $\bm{b}$.

This architecture is not an optimization but a fundamental necessity. It ensures that the enhanced intelligence of Rich-ARQ does not compromise the bedrock requirement of wireless communication: reliable and timely transmission.

In summary, our Rich-ARQ prototype embodies the successful translation of the AFC into a functional, standards-compatible system. It bridges the gap between theoretical design and practical deployment, providing tangible evidence that rich, neural-coded feedback is not just viable but operable within real-world constraints. We now proceed to evaluate its performance, quantifying the gains this new paradigm delivers over conventional feedback mechanisms.

\begin{table}[t]
	\caption{Configuration of evaluated schemes: $K$ is the number of information bits, $N$ is the maximum blocklength, and $R$ is the minimum mother code rate.}
    \centering
    \setlength{\tabcolsep}{4mm} 
    \begin{tabular}{ccccc}
        \toprule
        Codes & $K$ & $N$ & $R$\\
        \midrule
        Turbo-HARQ-CC & 47 & 144 & 0.326\\
        Polar-HARQ-CC & 47 & 144 & 0.326\\
        GBAF & 51 & 153 &0.333\\
        Rich-ARQ & 48 & 144 & 0.333\\
        Rich-ARQ (Light) & 48 & 144 & 0.333\\
        \bottomrule
	\end{tabular}
    \label{table_coding_config}
\end{table}

\section{Experimental Evaluation and Analysis}
\label{sec:experiments}
In this section, we turn to empirical validation to answer the question: does the proposed framework deliver on its promises in practice? 
Our evaluation moves from controlled simulations to over-the-air tests on our full-stack prototype, aiming to quantify its performance gains in reliability, latency, and practical feasibility. The experiments are structured to first validate the core advantage of Rich-ARQ over conventional 1-bit feedback schemes, then evaluate the latency advantages resulting from asynchronous design and lightweight model compared with existing DL-based feedback codes. Our findings demonstrate that Rich-ARQ is a practical and high-performance framework for real-world deployment.

To establish a fair comparison, we benchmark Rich-ARQ against three baseline schemes configured for a similar minimum mother code rate of $R \approx 1/3$:
\begin{itemize}[leftmargin=0.5cm]
    \item Turbo-HARQ-CC: A Turbo code with Chase Combining HARQ.
    \item Polar-HARQ-CC: A Polar code with Chase Combining HARQ.
    \item GBAF: A state-of-the-art DL-based feedback code, trained at a fixed $0$ dB forward SNR.
\end{itemize}

Detailed configurations are provided in Table~\ref{table_coding_config}, where $K$ is the number of information bits, $N$ is the maximum blocklength, $R$ is the minimum mother code rate, and ``Rich-ARQ (Light)'' refers to Rich-ARQ with the lightweight version of AFC.

\subsection{Rich-ARQ versus Conventional HARQ}
The foundational promise of Rich-ARQ is that replacing the minimalist 1-bit ACK/NACK with rich, multi-dimensional feedback creates a far more powerful collaboration between the decoder and encoder. We first validate this core proposition by examining the most direct metric of link reliability: the Packet Error Rate (PER). If the rich feedback indeed conveys more actionable information than a mere success/failure indication, it should translate into a substantial reduction in PER for a given channel condition.

\begin{figure}[t]
    \centering
    \includegraphics[width=0.4\textwidth]{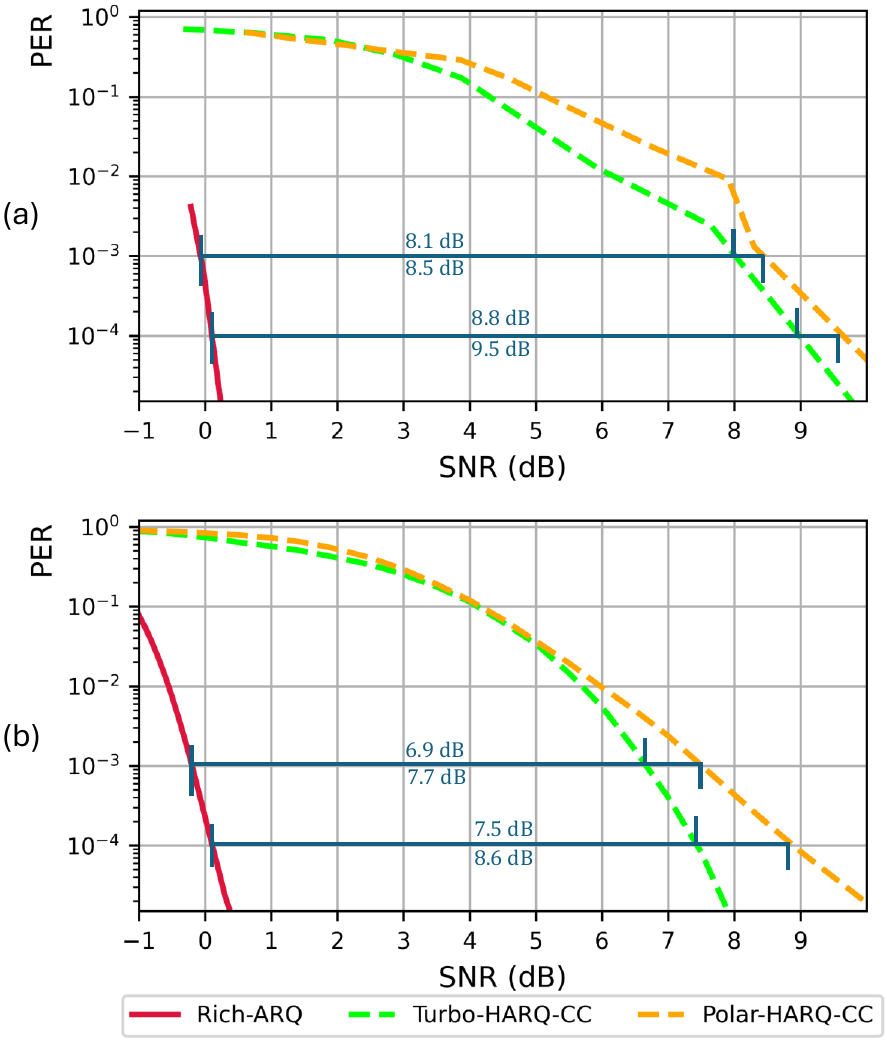}
    \caption{PER performance comparison among Rich-ARQ, Turbo-HARQ-CC, and Polar-HARQ-CC: (a) Over-the-air experimental results. (b) Complementary simulation results confirming the trend.}
    \label{fig_HARQ}
\end{figure}

\subsubsection{PER performance}
We deploy Turbo-HARQ-CC, Polar-HARQ-CC, and Rich-ARQ on our testbed under identical conditions. The over-the-air results, shown in Fig.~\ref{fig_HARQ}(a), confirm a dramatic advantage for Rich-ARQ. At a target PER of $10^{-4}$, it provides an $8.8$ dB and $9.5$ dB gain over Turbo-HARQ-CC and Polar-HARQ-CC, respectively. This significant margin persists across the SNR range. To isolate the gain attributable to the coding strategy itself from potential implementation-specific artifacts, we complement these experiments with simulations under perfect channel knowledge. The simulation results in Fig.~\ref{fig_HARQ}(b) robustly confirm the experimental trend, solidifying the conclusion that the rich feedback mechanism fundamentally enables more efficient information reconstruction.

\subsubsection{Coverage range analysis}
To quantify how the link-layer PER improvement translates into enhanced coverage, we analyze the maximum communication distance enabled by each coding scheme. The key parameter is the receiver sensitivity $S_{\text{rx}}$, defined as the minimum received signal power required to achieve a target PER.
A coding scheme operating at a lower SNR for the same PER (as observed in Fig.~\ref{fig_HARQ}) exhibits better (numerically lower) sensitivity, enabling it to tolerate greater path loss for a fixed transmit power.

We adopt the widely used log-distance path loss model to characterize large-scale fading:
$$\text{PL}(d) = \text{PL}_0 + 10 n \log_{10}\left(\frac{d}{d_0}\right),$$ 
where $d$ is physical distance between the transmitter and receiver, $\text{PL}(d)$ is the path loss (in dB) at distance $d$, $\text{PL}_0$ is the reference path loss at distance $d_0$, and $n=3$ is the path loss exponent typical for urban environments.

The maximum distance $d_{\max}$ is reached when the received power equals the receiver sensitivity $S_{rx}$.
For a system with fixed transmit power $P_{\text{tx}}$ (dBm) and antenna gains $G_{\text{tx}}$ and $G_{\text{rx}}$ (dBi), the maximum allowable path loss $\text{PL}_{\max}$ is
\begin{equation}
    \text{PL}_{\max} = P_{tx}+G_{tx}+G_{rx}-S_{rx},
\end{equation}

Setting $\text{PL}(d_{\max}) = \text{PL}_{\max}$ and solving for $d_{\max}$ yields the relationship between the sensitivities and coverage distances:
\begin{equation}
d_{\max} = d_0 \cdot 10^{\frac{\text{PL}_{\max} - \text{PL}_0}{10n}}.
\label{eq_d_max_corrected}
\end{equation}

To compare relative coverage distances of any two schemes 1 and 2, we analyze the distance ratio between them. From \eqref{eq_d_max_corrected}, we have
\begin{equation}
\frac{d_{\max,1}}{d_{\max,2}} = 10^{\frac{S_{\text{rx},2} - S_{\text{rx},1}}{10n}},
\end{equation}
where $S_{\text{rx},1}$ and $S_{\text{rx},2}$ are the receiver sensitivities (in dBm) for schemes 1 and 2, respectively, at the same target PER. For a fixed noise power spectral density $N_0$, the sensitivity difference in dB equals the SNR difference in dB: $S_{\text{rx},2} - S_{\text{rx},1} = \text{SNR}_2 - \text{SNR}_1$.

\begin{figure}[t]
    \centering
    \includegraphics[width=0.35\textwidth]{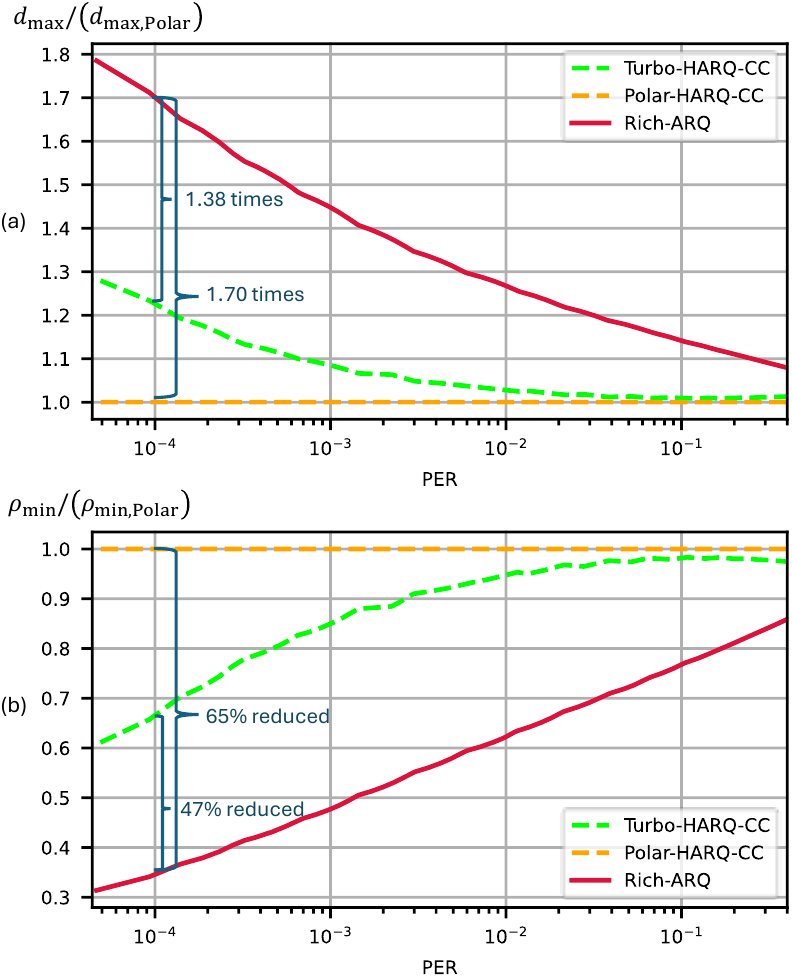}
    \caption{Coverage performance of Rich-ARQ: (a) Maximum communication distance normalized to the Polar-HARQ-CC baseline; (b) Minimum AP density normalized to the Polar-HARQ-CC baseline.}
    \label{fig_prac}
\end{figure}

From the PER-SNR characteristics in Fig.~\ref{fig_HARQ}(b) at a target PER of $10^{-4}$, Rich-ARQ operates at approximately 7.5 dB and 8.6 dB lower SNR than Turbo-HARQ-CC and Polar-HARQ-CC, respectively. These SNR advantages translate directly into sensitivity improvements. Assuming $n=3$, we calculate the relative coverage distances:
\begin{itemize}[leftmargin=0.5cm]
    \item Compared to Turbo-HARQ-CC: $\frac{d_{\max,\text{Rich-ARQ}}}{d_{\max,\text{Turbo}}} \approx 1.38$,
    \item Compared to Polar-HARQ-CC: $\frac{d_{\max,\text{Rich-ARQ}}}{d_{\max,\text{Polar}}} \approx 1.70$.
\end{itemize}

Thus, Rich-ARQ provides approximately $1.38\times$ and $1.70\times$ increases in maximum communication distance relative to Turbo-HARQ-CC and Polar-HARQ-CC, respectively.

The expanded coverage reduces the required infrastructure (i.e., AP) density for area coverage. Assuming a hexagonal cell layout with cell radius proportional to $d_{\max}$, the cell area $A$ scales as $A \propto d_{\max}^2$. Therefore, the minimum AP density $\rho_{\min}$ (APs per unit area) scales inversely with the square of the maximum distance: $\rho_{\min} \propto \frac{1}{d_{\max}^2}$.
Normalizing by Polar-HARQ-CC ($\rho_{\min,\text{Polar}} \propto 1$), we obtain:
\begin{itemize}[leftmargin=0.5cm]
    \item Rich-ARQ: $\rho_{\min,\text{Rich-ARQ}} \propto  0.35$,
    \item Turbo-HARQ-CC: $\rho_{\min,\text{Turbo}} \propto  0.66$.
\end{itemize}

This corresponds to approximately $47\%$ reduction in AP density compared to Turbo-HARQ-CC and $65\%$ reduction compared to Polar-HARQ-CC. These results demonstrate that Rich-ARQ can significantly extend cell coverage or, equivalently, substantially reduce infrastructure deployment costs while maintaining the same target reliability.

\subsection{Rich-ARQ versus GBAF}
Having established the superiority of Rich-ARQ over conventional HARQ, we now evaluate it against the state-of-the-art in DL-based feedback codes: the GBAF code. This comparison aims to demonstrate that Rich-ARQ is not merely an incremental improvement but a solution to the fundamental practical limitations that hinder prior feedback code deployments.

\subsubsection{Robustness to varying SNR}
A core limitation of existing feedback codes like GBAF is their optimization for a single, predetermined SNR, which fails to generalize to the time-varying channels characteristic of real deployments. We test this critical aspect by evaluating both Rich-ARQ and GBAF on our testbed across a range of SNRs.

\begin{figure}[t]
    \centering
    \includegraphics[width=0.37\textwidth]{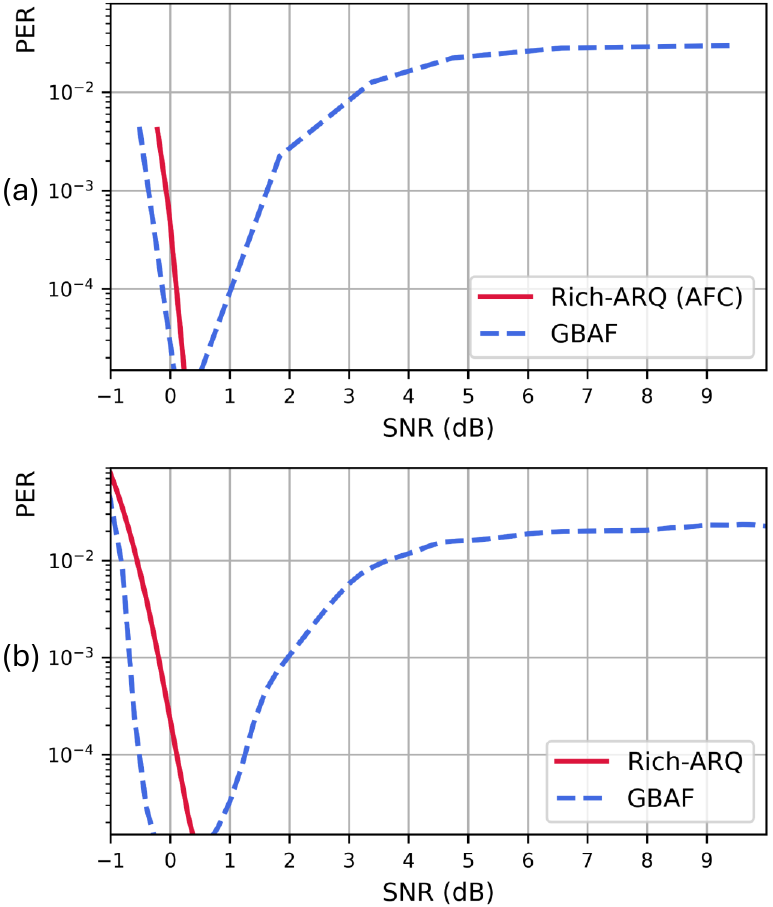}
    \caption{Performance comparison between Rich-ARQ and GBAF: (a) PER from over-the-air experiments; (b) PER from simulations. Rich-ARQ maintains robust performance across SNRs, while GBAF degrades rapidly away from its training point (0 dB).}
    \label{fig_GBAF}
\end{figure}

Fig.~\ref{fig_GBAF}(a) presents the over-the-air experimental results. While GBAF performs adequately near its trained SNR (0 dB), its PER degrades significantly outside this narrow operating point. This sharp decline indicates overfitting to a specific channel state and a lack of generalization, rendering it unsuitable for practical environments with SNR variation. In contrast, Rich-ARQ demonstrates robust performance, with a monotonic decrease in PER across the entire SNR range. This robustness stems from the model's learned ability to dynamically adapt its coding and feedback strategy using the SNR embedding $\gamma^{(t)}_{\text{emb}}$. These results confirm that Rich-ARQ effectively solves the SNR adaptability problem inherent in prior schemes.
The simulation results in Fig.~\ref{fig_GBAF}(b) corroborate these experimental findings, showing the same trend of superior and stable performance for Rich-ARQ under varying channel conditions.

\subsubsection{Latency reduction}
A key practical advantage of Rich-ARQ is its ability to break the fundamental latency bottleneck inherent in synchronous feedback protocols. Here, we provide a quantitative analysis of this latency reduction, assuming idealized conditions without uplink and downlink scheduling delays. The analysis focuses on the core processing and transmission latencies that are stable under fixed hardware and physical-layer settings.

We model the end-to-end latency for completing a session of $T$ transmission rounds. Three primary time components are defined for each round $t$:
\begin{itemize}[leftmargin=0.5cm]
    \item $\tau_{\text{enc}}$: The time required for the UE to generate a parity block $\bm{c}^{(t)}$ (encoding computation).
    \item $\tau_{\text{tx}}$: The time required to transmit $\bm{c}^{(t)}$ over the air from the UE to the AP.
    \item $\tau_{\text{fb}}$: The time required for the AP to generate the feedback packet $\widetilde{\bm{c}}^{(t)}$, and feed it back to the UE.
\end{itemize}

We combine encoding and transmission into a single forward interval $\delta=\tau_{\text{enc}}+\tau_{\text{tx}}$, and denote the feedback interval as $\widetilde{\delta}=\tau_{\text{fb}}$. 
In a synchronous scheme (e.g., GBAF), the encoder for round $t+1$ cannot start until the feedback for round $t$ is fully received and processed. This creates a sequential dependency where the forward transmission of round $t+1$ must wait for the completion of the feedback for round $t$. Consequently, the total latency to finish $T$ rounds is the sum of the forward intervals and the interleaved feedback intervals:
\begin{equation}
D_{\text{sync}} = T \cdot \delta + (T-1) \cdot \widetilde{\delta}.
\label{eq_sync_latency}
\end{equation}
In our prototype, typical values measured are $\delta=10$ ms (dominated by DNN encoding) and $\widetilde{\delta}=4$ ms. For $T=9$ rounds, this yields $D_{\text{sync}}=122$ ms. The forward processing and transmission time $\delta$ constitutes the dominant component (approximately $73.8\%$ of the total).

The asynchronous design of AFC fundamentally changes this timeline. The encoder is not blocked by the most recent feedback. In our implementation, when generating the codeword for round $t$, the encoder utilizes the feedback received up to round $t'=t-2$. This means the encoding computation for round $t$ can start as soon as the feedback for round $t-2$ arrives, overlapping with the round-trip time of rounds $t-1$ and the encoding and transmission time of round $t$.

To quantify the gain, we first examine the interval between the end of one feedback transmissions and the start of the next feedback transmission, denoted by $\delta^\prime$. In the steady state of the asynchronous pipeline, the time from the start of encoding for round $t$ to the start of encoding for round $t+1$ must accommodate:
(i) the interval between the end of the $(t-2)$-th feedback and start of the $(t-1)$-th feedback, i.e., $\delta^\prime$,
(ii) the generation and transmission of feedback for round $t-1$, i.e., $\widetilde{\delta}$, and
(iii) the interval between the end of the $(t-1)$-th feedback and the start of the $t$-th feedback, i.e., $\delta^\prime$.
In other words, the total time $\delta$ that be used for the $t$-th round encoding satisfies
\begin{equation}
\delta=2\delta^\prime + \widetilde{\delta}.
\end{equation}
Solving for $\delta^\prime$ yields:
\begin{equation}
\delta^\prime = \max\left(\frac{\delta - \widetilde{\delta}}{2},1\right),
\label{eq_delta_prime_derivation}
\end{equation}
where we introduce the maximization operation because the forward transmission consumes at least $1$ ms. 
Overall, the total latency of asynchronous coding can be written as
\begin{equation}
D_{\text{async}} = \delta + (T-1) \cdot \widetilde{\delta}+T \cdot \delta^\prime.
\label{eq_async_latency}
\end{equation}
Substituting the measured values $\delta=10$ ms and $\widetilde{\delta}=4$ ms, and $T=9$, we get $D_{\text{async}}=69$ ms.
Compared to the synchronous baseline of $122$ ms, Rich-ARQ achieves a latency reduction of $43.4\%$.

\begin{figure}[t]
    \centering
    \includegraphics[width=0.45\textwidth]{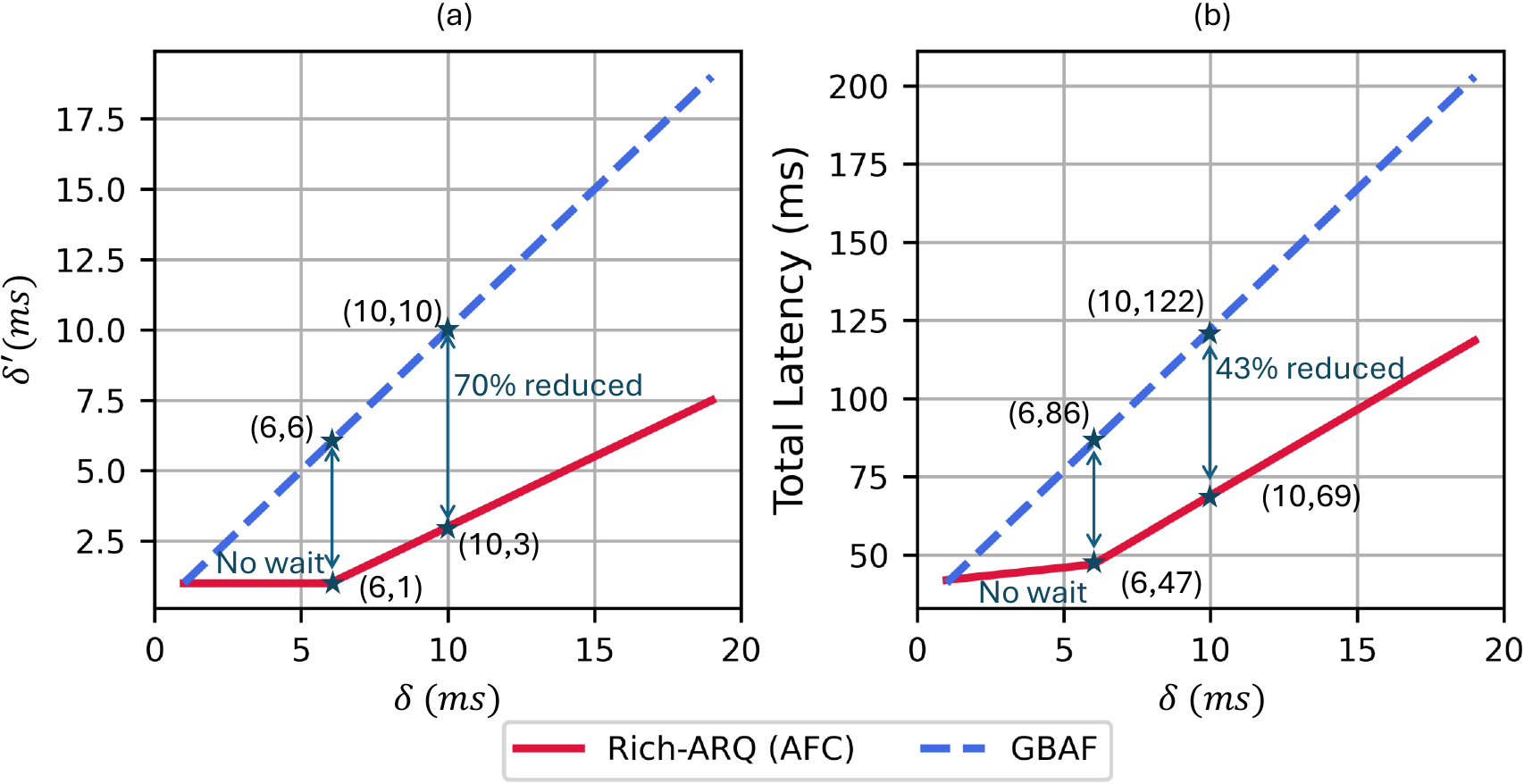}
    \caption{Effective forward interval $\delta^\prime$ and total latency as functions of forward interval $\delta$ (with fixed $\widetilde{\delta}=4$ ms): (a) $\delta^\prime$; (b) total latency. The asynchronous scheme maintains a lower $\delta^\prime$ and significantly reduces total latency.}
    \label{fig_latency_with_fwd}
\end{figure}

\begin{figure}[t]
    \centering
    \includegraphics[width=0.45\textwidth]{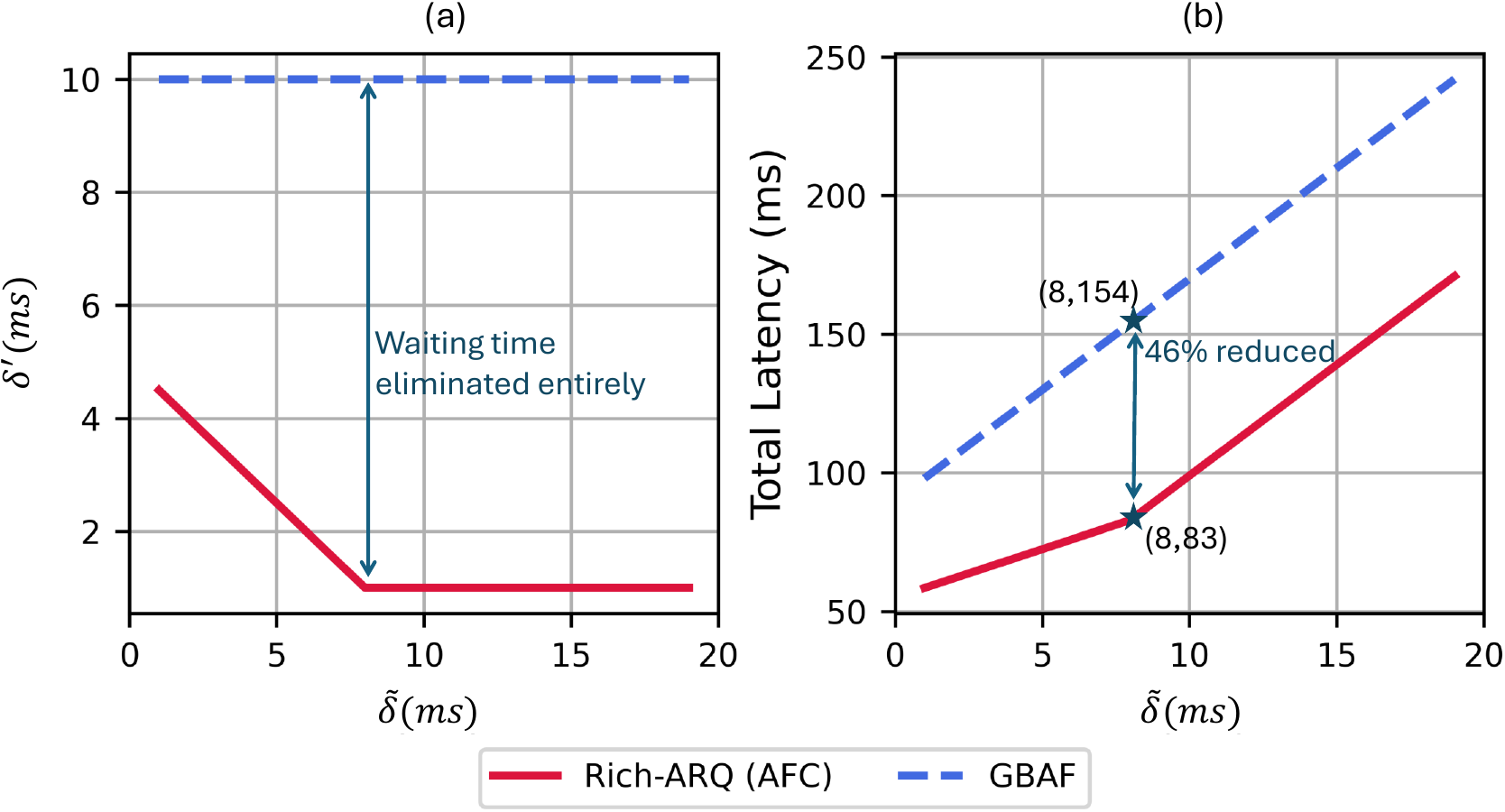}
    \caption{Effective forward interval $\delta^\prime$ and total latency as functions of feedback interval $\widetilde{\delta}$ (with fixed $\delta=10$ ms): (a) $\delta^\prime$; (b) total latency. As $\widetilde{\delta}$ increases, the asynchronous advantage persists.}
    \label{fig_latency_with_fbk}
\end{figure}

The latency reduction advantage of our system is generalizable across different operating conditions. In Fig.~\ref{fig_latency_with_fwd}, we examine how the performance of AFC varies with increasing forward interval $\delta$, while keeping the feedback interval fixed at $\widetilde{\delta} = 4$ ms. As shown in Fig.~\ref{fig_latency_with_fwd}(a), the interval $\delta'$ under asynchronous coding remains at $1$ ms for small values of $\delta$. This indicates that the encoder does not incur additional waiting time, and the remaining $1$ ms corresponds essentially to the unavoidable signal transmission time from transmitter to receiver. As $\delta$ increases further, $\delta'$ grows gradually, but its growth rate remains consistently lower than that of the synchronous scheme. At $\delta = 10$ ms, a typical value in our prototype, $\delta'$ is only $3$ ms, representing a $70\%$ reduction compared to the synchronous case. This demonstrates that if the forward encoding and transmission time is sufficiently small, it can be completely hidden within the feedback period; otherwise, asynchronous coding still achieves significant latency reduction. The impact on total latency follows a similar trend, as shown in Fig.~\ref{fig_latency_with_fwd}(b). Initially, the total latency of asynchronous coding increases slowly as $\delta$ increases, because most codewords' encoding time (from $\bm{c}^{(1)}$ to $\bm{c}^{(T-1)}$) is absorbed within the pipeline. Beyond this point, total latency increases but at a slower rate than that of synchronous coding.

In Fig.~\ref{fig_latency_with_fbk}, we evaluate the effect of increasing feedback interval $\widetilde{\delta}$, with $\delta$ fixed at $10$ ms. As $\widetilde{\delta}$ grows, $\delta'$ gradually decreases, as shown in Fig.~\ref{fig_latency_with_fbk}(a). This occurs because a longer feedback period provides more time to overlap with forward encoding, thereby reducing the effective forward interval. Eventually, $\delta'$ approaches $1$ ms, limited only by the physical transmission time. At this point, the system operates with minimal stall, and the encoder seldom needs to wait. The total latency, shown in Fig.~\ref{fig_latency_with_fbk}(b), increases with $\widetilde{\delta}$ but remains lower than that of synchronous coding across the range. For example, when $\widetilde{\delta} = 8$ ms, asynchronous coding reduces total latency by $46\%$ compared to the synchronous baseline. Although both schemes exhibit similar growth trends for larger $\widetilde{\delta}$, the absolute latency of asynchronous coding stays consistently lower, underscoring its robustness under varying feedback delays.

\subsection{Complexity and Feasibility of Lightweight Encoder}
\label{subsec:lightweight}

\begin{figure}[t]
    \centering
    \includegraphics[width=0.45\textwidth]{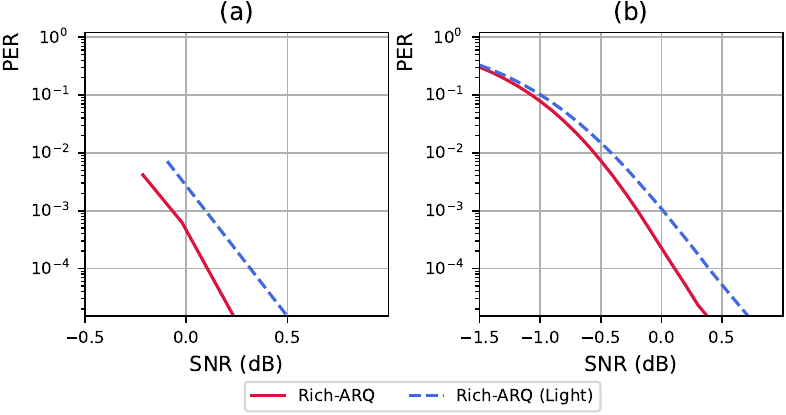}
    \caption{PER comparison between the full Rich-ARQ model and its lightweight version: (a) over-the-air experimental results; (b) simulation results under AWGN channels.}
    \label{fig_Light}
\end{figure}

To evaluate the efficiency of the proposed lightweight AFC encoder, we compare its complexity and performance with the full model. Table~\ref{table_light} summarizes the key metrics. The lightweight model reduces the number of parameters from 40.3K to 21.2K, a reduction of $47.4\%$. The computational cost, measured in floating-point operations (FLOPs), decreases from $661$K to $444$K (a $32.8\%$ reduction). In terms of practical runtime, the processing latency per forward pass (measured with a batch size of 8192 on an NVIDIA RTX 4090 GPU) drops from $111.07$ ms to $67.52$ ms, yielding a speedup of $39.2\%$. As shown in Fig.~\ref{fig_Light}, this significant complexity reduction incurs only a marginal performance degradation in both experimental and simulated settings. These results demonstrate that the lightweight model successfully preserves the core benefits of Rich-ARQ while becoming far more amenable to deployment on resource-constrained devices.

\begin{table}[t]
	\caption{Complexity comparison between the full and lightweight AFC models.}
    \centering
    \begin{tabular}{cccc}
        \toprule
        \thead{Model} & \thead{Params\\(K)} & \thead{FLOPs\\(K)} & \thead{Latency\\(ms/batch)}\\
        \midrule
        Full AFC Encoder & 40.3 & 661 &111.07\\
        Light AFC Encoder & 21.2 & 444 & 67.52 \\
        \bottomrule
	\end{tabular}
    \label{table_light}
\end{table}

While the asynchronous coding framework and the lightweight model already reduce latency and computational burden, a production-grade deployment would typically migrate the encoder to dedicated hardware (e.g., an FPGA or ASIC) to achieve real-time operation. Our current prototype runs on a general-purpose CPU/GPU platform, which introduces non-negligible data-synchronization overhead between the RF front-end and host memory, as well as limited neural-network inference efficiency. Deploying the lightweight AFC encoder on an FPGA can eliminate these software overheads and exploit massive parallelism.

To estimate the achievable performance on embedded platforms, we analyze the theoretical encoding latency based on the computational budget of the lightweight model ($444$K FLOPs). Since there is no unified standard for comparing neural-network inference across FPGA devices, we use the theoretical peak computational throughput, derived from the number of digital signal processors (DSPs) and their operating frequency, as a common basis. Assuming each DSP can perform one multiply-accumulate (MAC) operation per cycle and each MAC counts as two floating-point operations (FLOPs), the peak throughput in GFLOP/s is estimated as 
$2\times \text{DSP}~\text{count}\times \text{frequency}$ (GHz). Table~\ref{tab_fpga_perf} lists representative FPGA families and their corresponding peak throughput.

Given the peak throughput, the time $t_{\text{enc}}$ required to execute the $444$K FLOPs of the lightweight encoder can be estimated.
For instance, a Kintex-7 FPGA with a peak throughput of $5.69$ TFLOP/s would yield $t_{\text{enc}}\approx 0.078$ ms. Even a modest Spartan-7 device could complete the encoding in about $0.56$ ms. These values are well below the typical subframe duration ($1$ ms) in 4G/5G systems, indicating that the AFC encoding step can be comfortably hidden within the existing radio timeline.

\begin{table}[t]
\centering
\caption{Theoretical peak throughput and estimated encoding latency for selected FPGA families (assuming the lightweight AFC model with $444$K FLOPs).}
\begin{tabular}{lcccc}
\toprule
\thead{FPGA Family} & \thead{DSP\\(GMAC/s)} & \thead{Peak Throughput\\(GFLOP/s)} & \thead{Encoding Latency\\(GOP/s)}  \\
\midrule
Spartan-7 & 176   & 352  & 0.80  \\
Artix-7   & 929  & 1858  & 4.19  \\
Kintex-7  & 2,845   & 5690  & 12.80  \\
Virtex-7  & 5,335   & 10670  & 35.33 \\
\bottomrule
\end{tabular}
\label{tab_fpga_perf}
\end{table}

\section{Conclusion}
This work presented Rich-ARQ, a framework that fundamentally reimagines feedback, transforming it from a passive, one-bit verdict into an active, high-dimensional collaboration between the transmitter and receiver. By integrating DL directly into the physical-layer coding loop, we demonstrated that feedback can evolve beyond a simple retransmission trigger to become a rich, informative signal that dynamically guides the encoder toward optimal reliability.

Our journey from concept to a functional prototype yields several critical insights:
\begin{itemize}[leftmargin=0.5cm]
    \item The shift from synchronous to asynchronous feedback coding is not merely an optimization but an architectural necessity for practical deployment. This design liberates the system from a fragile dependency on instantaneous feedback, creating a robust pipeline that significantly reduces end-to-end latency and gracefully accommodates the variable delays inherent in wireless networks.
    \item Robustness to time-varying channels must be inherent, not an afterthought. By co-designing the training curriculum with the stochastic nature of real SNR traces, the learned coding strategy becomes intrinsically adaptive, ensuring stable performance across a dynamic range of channel conditions, a requirement overlooked by prior neural feedback codes.
    \item Most importantly, we validate that this vision transcends simulation. The construction of a full-stack, standard-compliant software-defined radio prototype proves that neural-coded feedback can be successfully integrated into real-time wireless systems, turning theoretical advantages into tangible, over-the-air performance gains.
\end{itemize}

The design principles of Rich-ARQ are particularly well-suited for demanding applications such as mission-critical IoT, ultra-reliable communication, and multiple access channels with correlated sources, where its ability to deliver robust short-packet performance is paramount.

Looking forward, the impact of Rich-ARQ extends beyond a specific coding scheme or even the physical layer; it seeds a new paradigm for interactive communication. Future systems may leverage rich feedback to convey semantic or goal-oriented information, guiding transmissions to fulfill specific application-level objectives rather than merely ensuring bit-level reliability. Rich-ARQ provides both the foundational framework and a practical proof-of-concept for this emerging class of collaborative, feedback-driven communication protocols.

\appendices

\bibliographystyle{IEEEtran}
\bibliography{references}

\end{document}